\documentstyle{l-aa}
\input psfig
\psfull
\def\be{\begin{equation}}
\def\ee{\end{equation}}
\def\ba{\begin{eqnarray}}
\def\ea{\end{eqnarray}}
\def\wisk#1 {\ifmmode{#1}\else{$#1$}\fi}
\def\kms    {\rm ~km~s^{-1}}
\def\ob    {\Omega_{b}}
\def\n      {n}

\def\ol    {\lambda_{o}}
\def\oo    {\Omega_{o}}

\def\oh    {\Omega_{HDM}}

\def\etal   {{\sl et al.}~\rm}
\def\apj    {{\rm Ap.J.$\!$}~\rm}

\begin{document}

\thesaurus{
02  % A&A Section 2: Cosmology
(12.03.1;      % {\it Cosmology)} cosmic microwave background
12.03.3;       % Cosmology: observations
12.03.4)        % {\it Cosmology)} theory
}
\title{Cosmic microwave background observations:
implications for Hubble's constant and the spectral parameters $\n$ and $Q$
in cold dark matter critical density universes}
% Yield Low Values for Hubble's Constant and Provide Limits on the Spectral Parameters $\n$ and $Q$}
\author{ Charles H. Lineweaver \inst{1}
\and Domingos Barbosa  \inst{1,2}
}
\institute{Observatoire astronomique de Strasbourg, U.L.P., 67000 Strasbourg, France. 
\and
Centro de Astrof\'{\i}sica da U.P., Rua do Campo Alegre 823, 4150 Porto, Portugal.}

\offprints{Charley Lineweaver, charley@cdsxb6.u-strasbg.fr}
\markboth{Cosmic Microwave Background Observations }{Implications for Hubble's Constant...}

\date{Received ;     accepted        }
\maketitle
\markboth{Cosmic Microwave Background Observations }{Implications for Hubble's Constant...}

%%******************************************************************
\begin{abstract}
Recent cosmic microwave background (CMB) measurements  over a large range of 
angular scales have become sensitive enough to provide interesting 
constraints on cosmological parameters within a restricted class of models.
We use the CMB measurements to study inflation-based,  cold dark matter (CDM)
critical density universes. 
We explore the 4-dimensional parameter space having as free parameters, Hubble's 
constant $H_{o}$, baryonic fraction $\ob$, 
the spectral slope of scalar perturbations $\n$ and the power spectrum 
quadrupole normalization $Q$.
We calculate $\chi^{2}$ minimization values and likelihood intervals for 
these parameters.
Within the models considered, a low value for the Hubble constant is preferred:
$H_{o}= 30^{+18}_{-7}\:\kms Mpc^{-1}$.
The baryonic fraction is not as well-constrained by the CMB data: 
$\ob = 0.07^{+ 0.24}_{-0.07}$.
The power spectrum slope is
$\n=0.91^{+0.20}_{-0.12}$.
The power spectrum normalization is
$Q=18 \pm 2.5\:\mu$K.
The error bars on each parameter are approximately $1\sigma$ and are 
for the case where the other 3 parameters have been marginalized.
If we condition on $n=1$ we obtain the normalization $Q= 17 \pm 1.0\:\mu$K.
The permitted regions of the 4-D parameter space are presented in a series
of 2-D projections.
In the context of the CDM critical density universes considered here, current CMB 
data favor a low value for the Hubble constant. Such low-$H_{o}$ models are consistent
with Big Bang nucleosynthesis, cluster baryonic fractions, the large-scale 
distribution of galaxies and the ages of globular clusters;  
although in disagreement with direct determinations of the Hubble constant.

\end{abstract}

%\clearpage
%%%%%%%%%%%%%%%%%%%%%%%%%%%%%%%%%%%%%%%%%%%%%%%%%%%%%%%%%%%%%%%%%%%
\section{Introduction}
\label{sec:intro}
%%%%%%%%%%%%%%%%%%%%%%%%%%%%%%%%%%%%%%%%%%%%%%%%%%%%%%%%%%%%%%%%%%%

The standard picture of structure formation relies on the gravitational 
amplification of initially small perturbations in the matter distribution.
The origin of these fluctuations is unclear, but a popular 
assumption is that these fluctuations originate in the very early universe
during an inflationary epoch. The most straightforward incarnation of
this inflationary scenario predicts that
the fluctuations are adiabatic, Gaussian, Harrison-Zeldovich ($n = 1$) and
that the  Universe is spatially flat (Kolb \& Turner 1990).
To avoid violating primordial nucleosynthesis constraints, the Universe should be 
dominated by non-baryonic matter. The cold dark matter (CDM) model has been the preferred
model in the inflationary scenario (Peebles 1982, Liddle and Lyth, 1993).

The statistical properties of CMB fluctuations provide an ideal tool 
for testing CDM models. CMB data offer valuable information not only on
the scenario of the origin of cosmic structures, but also on the early 
physics  of the Universe and the cosmological parameters that characterize the 
Universe.
Using the CMB to determine these parameters is the beginning of a 
new era in cosmology.
This truly cosmological method probes scales much larger and
epochs much earlier ($z > 1000$) than more traditional techniques 
which rely on supernovae, galaxies, galaxy clusters and other low-redshift objects.
The CMB probes the entire observable universe.

Acoustic oscillations of the baryon--photon fluid at 
recombination produce peaks and valleys in the CMB 
power spectrum at sub-degree angular scales.
Measurements of these model-dependent peaks and valleys
have the potential to determine many important 
cosmological parameters to the few percent level 
(Jungman \etal 1996, Zaldarriaga \etal 1997).
Within the next decade, increasingly accurate sub-degree scale
CMB observations from the ground, from balloons and particularly 
from two new satellites ( MAP: Wright \etal 1996, Planck Surveyor: 
Bersanelli \etal 1996)
will tell us the ultimate fate of the Universe ($\oo$),
what the Universe is made of ($\ob$, $\Omega_{CDM}$) and 
the age and size of the Universe ($h\equiv H_{o}/(100/\kms Mpc^{-1}$) 
with unprecedented precision.

In preparation for the increasingly fruitful harvests of data,
it is important to determine what the combined CMB data can {\it already}
tell us about the cosmological parameters.
In Lineweaver \etal (1997), (henceforth ``paper I''), we compared the 
most recent CMB data to predictions of COBE normalized flat 
universes with Harrison-Zel'dovich ($n=1$) power spectra.
We used predominantly goodness-of-fit statistics to locate the regions of 
parameter space preferred by the CMB data.
We explored the $h - \ob$ plane and  the $h - \ol$ plane.

In the present paper we focus on the range for $h$ favored by the CMB in the 
context of CDM critical density universes and
we broaden the scope of our exploration to
the 4-dimensional parameter space: $h$, $\ob$, $\n$ and $Q$.
Our motivation for choosing this 4-D subspace
of the higher dimensional parameter space is that (i) it 
is the largest dimensional subspace that we can explore at a 
reasonable resolution with the means available and  (ii) it is centered on
the simplest CDM model: $\oo = 1$, $\ol = 0$, $n=1$.
This model is arguably the simplest scenario for the formation of large-scale 
structure. One of our goals is to see what is required of such a model if it is 
to explain the current set of large-scale structure data, and what could 
eventually force us to accept the fine-tuning demanded by the inclusion of 
another cosmological parameter, such as the cosmological constant.

Hubble's constant $H_{o}$ is possibly the most important parameter in cosmology,
giving the expansion rate, age and size of the Universe.
Recent, direct, low-redshift measurements fall in the range [45-90] but may be 
subject to unidentified systematic errors.
Thus it is important to have  different methods which may not be subject to the same 
systematics. For example, CMB determinations of $h$ are distance-ladder-independent.
Current CMB data are not of high enough quality to draw definitive model-independent
conclusions, however in the restricted class of models considered here,
the CMB data are already able to provide interesting constraints. 

The quantity $\ob$ is important because we would 
like to know what the universe is made of and how much normal baryonic 
matter exists in it.  The combination $\ob h^2$ is relatively
well-constrained by the observations and theory of primordial nucleosynthesis, 
but the uncertainty on the Hubble constant means that the value
of $\ob$ is rather poorly constrained.
The question of just how many baryons there are in the Universe
has received close attention recently due to estimates
of the baryon fraction in galaxy clusters and attempts to constrain $\ob$
by measuring the deuterium in high-redshift quasar absorption systems.

The parameter $n$ is the primordial power spectrum slope
that remains equal to its primordial value at the largest scales (low $\ell$).
It is important because it's measurement is a glimpse
at the primordial universe.
Although generic inflation predicts 
$\n=1$, a larger set of plausible inflationary models is 
consistent with $ 0.7 \la  \n \la 1.0$.
Model power spectra and particularly the amplitude of the first
peak depend strongly on $\n$. 
Thus, an important limitation of paper I was the restriction to $\n=1$. 
By adding $\n$ as a free parameter we obtain observational limits on $\n$ 
and quantify the reduced  constraining ability of the CMB observations 
when $\n$ is marginalized.

The power spectrum quadrupole normalization $Q$ is important because it normalizes
all models. 
Here we treat $Q$ as a free parameter.
 
We examine how the contraints on any one of these parameters change 
as we condition on and marginalize over the other parameters.
We  obtain $\chi^{2}$ minimization values and likelihood intervals for
$h$, $\ob$, $\n$ and $Q$.
As in paper I, we take advantage of the recently available fast
Boltzmann code (Seljak and Zaldarriga 1996) to make a detailed
exploration of parameter space.  

All the results reported here were obtained under a restrictive set of
assumptions. We assume inflation-based CDM models of structure formation
with Gaussian adiabatic initial conditions in critical density universes ($\oo =1$) 
with no cosmological constant ($\ol = 0$).
We have ignored the possibility of early reionization and any gravity wave 
contribution to the spectra.
We do not test topological defect models.
We use no hot dark matter.
We have used the helium fraction $Y_{He}=0.24$ and a mean CMB 
temperture $T_{o}=2.728$ K.  
Although we have not yet looked carefully at how dependent our results are on 
these assumptions we make some informed estimates in 
Sect. \ref{sec:caveats} where we also discuss previous work
using similar data sets and similar methods to look at
different families of models.

In Sect. \ref{sec:method} we describe the data analysis.
In Sect. \ref{sec:horesults} we present our results for $h$ and $\ob$ and discuss 
their dependence on some plausible variations in the data analysis.
We discuss non-CMB constraints and compare them with our results in
Sect. \ref{sec:noncmbconstraints}. 
In Sect. \ref{sec:nqresults} we present our results for $n$ and $Q$.
In Sect. \ref{sec:summary} we add some caveats and summarize.

%%%%%%%%%%%%%%%%%%%%%%%%%%%%%%%%%%%%%%%%%%%%%%%%%%%%%%%%%%%%%%%%%%%
\section{Method}
\label{sec:method}
\subsection{Data}
The data used are described in paper I,
however we have updated some data points and now include several 
more measurements: \\
$\bullet$ updated Tenerife point (Hancock \etal 1997):
      $\delta T_{eff} = 34.1^{+15.5}_{-9.3}$ at $\ell_{eff} = 20$,\\
$\bullet$ added BAM point (Tucker \etal 1997):
$\delta T_{eff} = 55.6^{+29.6}_{-15.2}$ at $\ell_{eff} = 74$,\\
$\bullet$ updated the two Python points (Platt \etal 1997):
$\delta T_{eff} = 60^{+15}_{-13}$ at $\ell_{eff} = 87$
and $\delta T_{eff} = 66^{+17}_{-16}$ at $\ell_{eff} = 170$,\\
$\bullet$ added MSAM single- and double-difference
points (Cheng \etal 1996)
$\delta T_{eff} = 40.7^{+30.5}_{-17.0}$ at $\ell_{eff} = 159$
$\delta T_{eff} = 44.4^{+23.9}_{-14.5}$ at $\ell_{eff} = 263$.

\noindent With the exception of the Saskatoon points, the calibration
uncertainties of the experiments were added in quadrature to the
error bars on the points.
The MSAM values are the weighted averages of the first and 
second MSAM flights. The error bars assigned encompass the $1\sigma$ 
limits from both flights.
In paper I, we did not include the MSAM points because of possible 
correlations with the Saskatoon results.  However the MSAM 
results are substantially lower than the Saskatoon results in this crucial
high-$\ell$ region of the power spectrum and it is not clear
that avoiding such correlations is more important than the additional
information provided by the MSAM data.
The figures presented in this paper include the MSAM points however we have 
also performed $\chi^{2}$ calculations without MSAM. 
We discuss these results in Sect. \ref{sec:sk}.

\subsection{Calculation}
\label{sec:chi}
A two-dimensional version of our $\chi^{2}$ calculation is described 
in paper I. In this work we generalize to 5 dimensions and use a likelihood
approach to determine the parameter ranges.
We treat the correlated calibration uncertainty of the 5 Saskatoon points as a 
nuisance parameter ``$u$''.  
For each point in 5-D space we obtain a value for 
$\chi^{2}( h, \ob, \n, Q, u)$. 
We assume $u$ comes from a Gaussion distribution about 
its nominal value with a dispersion of $14\%$. This Gaussian assumption amounts to 
adding $[(1-u)/0.14]^{2}$ to the $\chi^{2}$ calculation described in paper I.
For example, in paper I, our notation Sk-14, Sk0 and Sk+14 corresponds to 
$u= (0.86, 1.00, 1.14)$ respectively  (however we did not
assume a Gaussian distribution and so did not add the extra factor to the 
$\chi^{2}$ values).

For each point in the 4-D space of interesting parameters, $u$ takes on the value 
which minimizes the $\chi^{2}$ (Avni 1976, Wright 1994).
At the minimum in 4-D, $\chi^{2}_{min}$, the parameter 
values are the best-fit parameters. To obtain error bars on 
these values, we determine the 4-D surfaces which satisfy
$\chi^{2}(h, \ob, \n, Q) = \chi^{2}_{min} +x$ where $x = [ 1,\; 4,\; 9,\; 16]$.
Under the assumption that the errors on the data 
points are Gaussian (cf. de Bernardis \etal 1996), the $x=1$  ellipsoid can be
projected onto any of the axes to get the
$1\sigma$ confidence interval for the parameter of that axis.
If one is not interested in 1-D intervals but rather in
the confidence regions for 2 parameters simultaneously, then one would use
$x = [ 2.3,\; 6.17,\; 11.8,\; 19.3]$ (see Press \etal 1989 for details).
To make the figures, we project the 4-D surfaces onto the two dimensions of 
our choice and obtain contours which we project onto either axis.

We have normalized the power spectra using the conversion
$ C_{\ell} = Q^{2}\frac{4\pi}{5} \frac{C^{\prime}_{\ell}}{C^{\prime}_{2}}$,
where $C^{\prime}_{\ell}$ is the power spectrum output of the Boltzmann code.

%%%%%%%%%%%%%%%%%%%%%%%%%%%%%%%%%%%%%%%%%%%%%%%%%%%%%%%%%%%%%%%%%%%
%\clearpage
\begin{figure}[htbp]
%\picplace{5cm}
\centerline{\psfig{figure=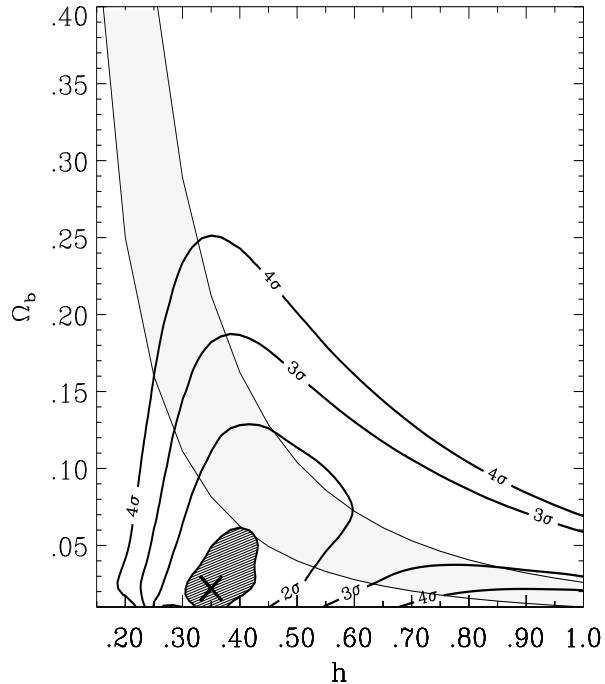,height=10.0cm,width=9.cm,bbllx=10pt,bblly=120pt,bburx=594pt,bbury=690pt}}
%\centerline{\psfig{figure=hob1.ps,height=10.0cm,width=9.cm,bbllx=10pt,bblly=120pt,bburx=594pt,bbury=690pt}}
%%%BoundingBox: 28 70 594 636
%
\caption{Likelihood contours in the $h - \ob$ plane from recent CMB measurements.
We have conditioned on $\n=1$ and $Q=17\; \mu$K.
The `{\bf X}' marks the minimum.
The contours are at levels $\chi^{2}_{min} + x$ where $x = [ 1,\; 4, \;9, 16]$
(see Section \protect\ref{sec:chi}). 
When projected onto either of the axes these regions give
the approximate size of the 1, 2, 3 and $4\sigma$ confidence intervals 
respectively.
The area within the $1\sigma$ contour has been shaded dark grey.
The result, $h = 0.35^{+0.08}_{-0.05}$, $\ob = 0.02^{+0.04}_{-0.02}$, is given in
Table I.
Big Bang nucleosynthesis predictions favor the light grey band defined by
$0.010 < \ob h^{2} < 0.026$ (see Sect. \protect\ref{sec:bbn}).}
\label{fig:hob1}
\end{figure}
%%%%%%%%%%%%%%%%%%%%%%%%%%%%%%%%%%%%%%%%%%%%%%%%%%%%%%%%%%%%%%%%%%%%

%%%%%%%%%%%%%%%%%%%%%%%%%%%%%%%%%%%%%%%%%%%%%%%%%%%%%%
\begin{figure}[htbp]
%\picplace{5cm}
\centerline{\psfig{figure=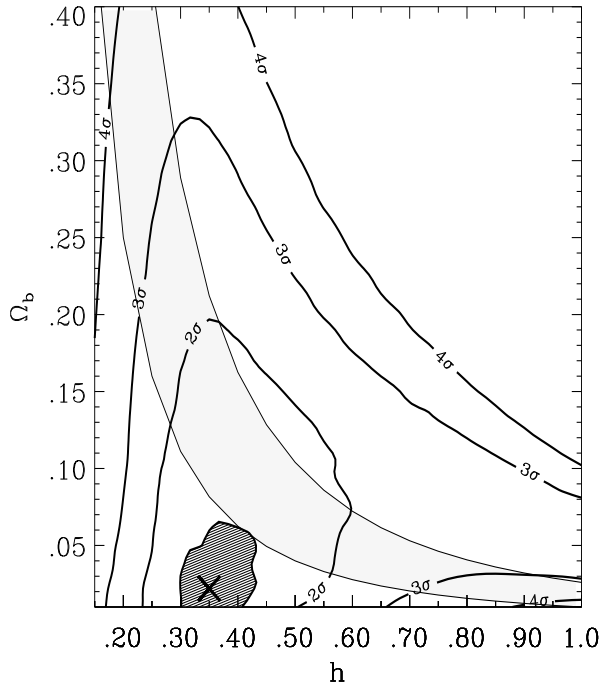,height=10.0cm,width=9.cm,bbllx=10pt,bblly=120pt,bburx=594pt,bbury=690pt}}
%\centerline{\psfig{figure=hob2qfree.ps,height=10.0cm,width=9.cm,bbllx=10pt,bblly=120pt,bburx=594pt,bbury=690pt}}
%\centerline{\psfig{figure=hob2.ps,height=9cm,width=\hsize,bbllx=10pt,bblly=120pt,bburx=594pt,bbury=690pt}}
%%%BoundingBox: 28 70 594 636
%
\caption{Contours and notation as in previous figure except here
the normalization $Q$ has become a free parameter, i.e.,
for a given $h$ and $\ob$, $Q$ takes on the value that minimizes the 
value of $\chi^{2}$ at that point. 
The minimum and the $1\sigma$ errors on $h$ and $\ob$ are the same as in 
Figure \protect\ref{fig:hob1} however the 2, 3, and $4\sigma$ contours 
are noticeably larger.
The higher values of $\ob$ permitted here correspond to $Q < 17\;\mu$K. 
}
\label{fig:hob2}
\end{figure}
%%%%%%%%%%%%%%%%%%%%%%%%%%%%%%%%%%%%%%%%%%%%%%%%%%%%%%%%%%%%%%%%%%%%

\section{Results for $h$ and $\ob$}
\label{sec:horesults}
\begin{figure}[htbp]
%\picplace{5cm}
\centerline{\psfig{figure=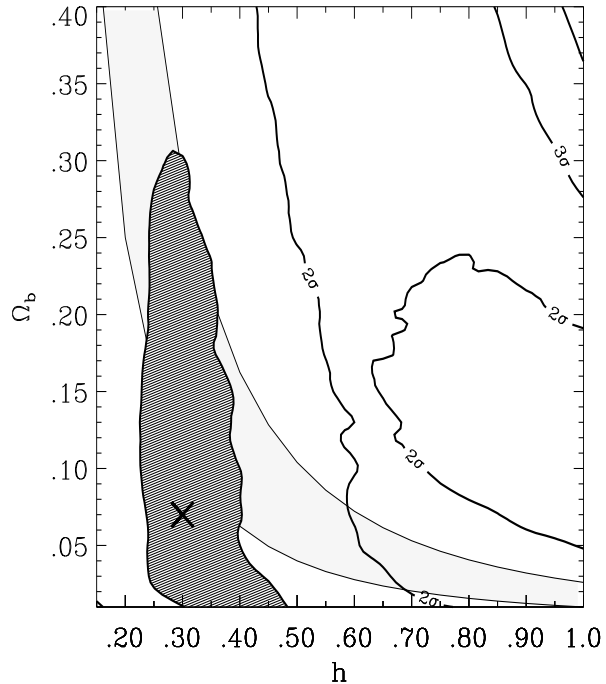,height=10.0cm,width=9.cm,bbllx=10pt,bblly=120pt,bburx=594pt,bbury=690pt}}
%\centerline{\psfig{figure=hob3i.ps,height=10.0cm,width=9.cm,bbllx=10pt,bblly=120pt,bburx=594pt,bbury=690pt}}
%\centerline{\psfig{figure=hob1.ps,height=9cm,width=\hsize,bbllx=10pt,bblly=120pt,bburx=594pt,bbury=690pt}}
%%%BoundingBox: 28 70 594 636
%
\caption{Same as previous figure except here both $Q$ {\it and} $n$ are
free parameters. The preferred value of $h$ stays low:
$h= 0.30^{+0.18}_{-0.07}$ but at $2\sigma$, $h$ can assume
all values tested, i.e., the projection of the $2\sigma$ region onto the $h$ axis covers the entire axis.
We also find $\ob = 0.07^{+0.24 }_{-0.07}$.
At the minimum $n=0.91$ and $Q = 18\;\mu$K.
In contrast to the previous figure the higher values of $\ob$ permitted correspond 
to $Q > 17\;\mu$K and $n < 1$. 
On the right, the high $h$, $\ob \sim 0.15$ region within the $2\sigma$ contour 
has $n \sim 0.7$ and $Q\sim 20\;\mu$K.
}
\label{fig:hob3}
\end{figure}

The permitted regions of the 4-D parameter space are presented in a series
of 2-D projections which contain likelihood contours
from a combination of the most recent CMB measurements.
There are four groups of figures corresponding to the four planes 
$h - \ob$, $h-n$, $h-Q$ and $n-Q$; Figures
\ref{fig:hob1} - \ref{fig:hob3}, \ref{fig:hn3},
\ref{fig:hq2obfree} - \ref{fig:hq3} and \ref{fig:nq1} - \ref{fig:nq3} 
respectively. 
The thick `{\bf X}' in each figure marks the minimum.
Areas within the $1\sigma$ contours have been shaded.

%%%%%%%%%%%%%%%%%%%%%%%%%%%%%%%%%%%%%%%%%%%%%%%%%%%
%%%%%%%%%%%%%%%%%%%%%%%%%%%%%%%%%%%%%%%%%%%%%%%%%%%%%%

%%%%%%%%%%%%%%%%%%%%%%%%%%%%%%%%%%%%%%%%%%%%%%%%%%%%%%
\begin{figure}[Htbp]
%\picplace{5cm}
%\centerline{\psfig{figure=hob5_nocmb.ps,height=10.0cm,width=9.cm,bbllx=10pt,bblly=120pt,bburx=594pt,bbury=690pt}}
%\centerline{\psfig{figure=hobnocmbnoage.ps,height=10.0cm,width=9.cm,bbllx=10pt,bblly=120pt,bburx=594pt,bbury=690pt}}
\centerline{\psfig{figure=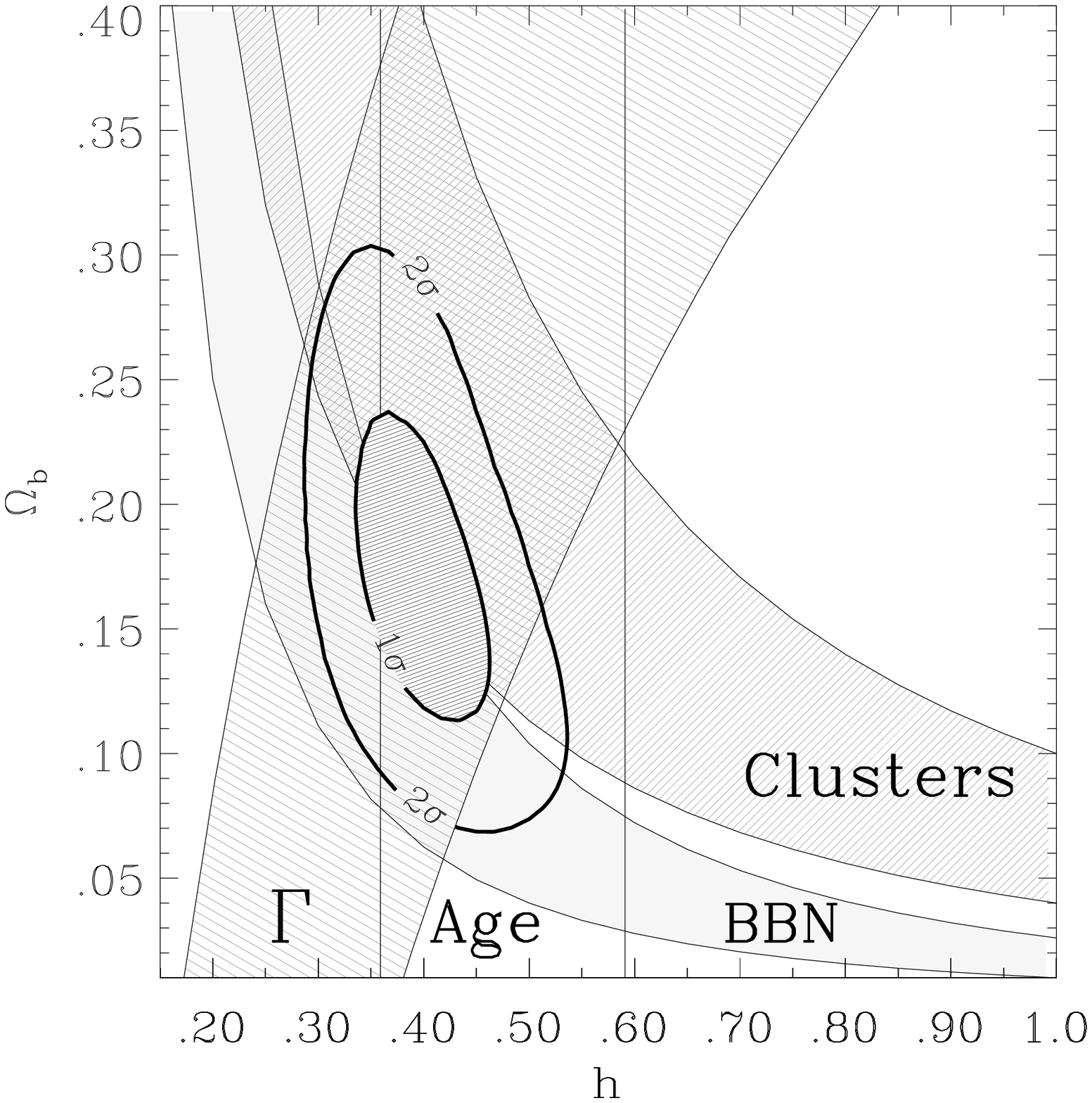,height=10.0cm,width=9.cm,bbllx=10pt,bblly=120pt,bburx=594pt,bbury=690pt}}
\caption{This plot has no CMB information in it.
The bands are the constraints from 4 non-CMB measurements
discussed in Sect. \protect\ref{sec:noncmbconstraints}:
Big Bang nucleosynthesis (``BBN''), cluster baryonic fraction  (``Clusters''),
the galaxy and cluster scale density fluctuation shape parameter (``$\Gamma$'')
and the age of the oldest stars in globular clusters 
(``Age'': region between the vertical lines).
To quantify the combination of 
these 4 constraints,
we show contours obtained by assuming for each constraint
a two-tailed Gaussian probability around the central values.
The 1 and $2\sigma$ contours are thus the result of an approximate joint 
likelihood of 4 non-CMB constraints for $\oo =1$, $\ol=0$ CDM models.
A low $h$ is preferred. This was pointed out in Bartlett \etal (1995).
The combination of these four, independent, non-CMB measurements yields
% three: $H_{o} \approx 36^{+8}_{-6}$ and $\ob = 0.19^{+0.06}_{-0.06}$.
$H_{o} \approx 40 \pm 7$ and $\ob \approx 0.18 \pm 0.06$.
This region of the $h-\ob$ plane is very similar to the region 
preferred by the CMB data (compare Figure \protect\ref{fig:hob3}).
The point of this diagram {\it is not} to show that $h$ is low since
we have ignored the numerous, more direct,  $z \sim 0$ measurements of $h$
which find $h \approx 0.70 \pm 0.10$.
The point of this diagram {\it is} to show that an important set of independent constraints 
also favors low values of $h$ in $\oo = 1$, $\ol=0$ CDM models.
}
\label{fig:hobnocmb}
\end{figure}
%%%%%%%%%%%%%%%%%%%%%%%%%%%%%%%%%%%%%%%%%%%%%%%%%%%%%%%%%%%%%%%%%%%%%%%%%%%%%%%%%%%%

\noindent The best-fit values and confidence intervals displayed in the 
figures are summarized in Table I which thus contains the main results of 
this work.
The values of $h$, $\ob$, $\n$ and $Q$ at the minimum of
the 4-D $\chi^{2}$ are given with
error bars from the projection onto 1-D of the 
$\chi^{2}_{min} + 1$ surface.
In Figures \ref{fig:hob1} through \ref{fig:hobnocmb} the region preferred by
Big Bang nucleosynthesis (BBN) 
is shaded light grey ($0.010 < \ob h^{2} < 0.026$, see Sect. \ref{sec:bbn}).

In Figure \ref{fig:hob1} we have conditioned on $\n=1$ and $Q=17\; \mu$K.
The results are $h = 0.35^{+0.08}_{-0.05}$, $\ob = 0.02^{+0.04}_{-0.02}$.
At $2\sigma$, $0.25 < h < 0.60$ and $0.0 < \ob < 0.13$.

The contours and notation of Figure \ref{fig:hob2} are the same as
in Figure \ref{fig:hob1} except that we have let the normalization $Q$
be a free parameter. That is, for each value of $h$ and $\ob$, $Q$ takes
on the value (within the discretely sampled range) which minimizes $\chi^{2}(h, \ob)$.
The minimum and the $1\sigma$ errors on $h$ and $\ob$ are the same as in 
Figure \protect\ref{fig:hob1}; $h$ stays low.
The 2, 3, and $4\sigma$ contours are noticeably larger than in 
Figure \ref{fig:hob1}.
Within the $2\sigma$ contours, the higher values of $\ob$ permitted correspond to $Q < 17\;\mu$K.

Figure \ref{fig:hob3} displays the main result for $h$ of this paper.
The result is more general than the results of Figures \ref{fig:hob1} and
\ref{fig:hob2} since no restrictions on $n$ and $Q$ are used.
Examining Figures \ref{fig:hob1}, \ref{fig:hob2} and  \ref{fig:hob3} sequentially,
the dark grey $1\sigma$ contours can be seen to get larger as we 
first condition on and then marginalize over $n$ and $Q$.
With both $n$ and $Q$ marginalized we obtain $h=0.30^{+0.18}_{-0.06}$
and $\ob = 0.07^{+0.24}_{-0.07}$ where the error bars are approximately 
$1\sigma$. At the minimum, $n=0.91$ and $Q=18\;\mu$K.
The $\chi^{2}$ value at this minimum is 16.
The number of degrees of freedom is 23 (=  27 data points - 
1 nuisance parameter - 3 marginalized parameters).
The probability of finding a $\chi^{2}$ value this low or lower is
$\sim 15\%$. Thus the fit obtained is ``good''.

The region of the $h- \ob$ plane acceptable to both the BBN and CMB have
low values of $h$.
Large values of $h$, especially in the BBN region (lower right of plot) are not favored 
by the CMB data. However at $2\sigma$, $h$ is unconstrained
since the projection of the $2\sigma$ contours onto the $h$
axis spans the entire axis.

Since $n$ is a free parameter the amplitude, but not the location, of the Doppler 
peak can vary substantially. 
This explains the shape of the 68\% region allowed by the data: the 
possible range of  $\Omega_b$ is enlarged, but not the confidence interval of the 
Hubble constant, which seems to be predominantly determined by the position of 
the Doppler peak.

The disjoint $2\sigma$ contour region on the right is
characterized by the parameter values 
$h \ga 0.65$, $\ob \sim 0.15$,
$n \approx 0.7 \pm 0.1$ and $Q \approx 20 \pm 2\:\mu$K.
The minimum $\chi^{2}$ of this region is at $h=1.0$.

\subsection{Robustness of results to data analysis choices}
\label{sec:sk}

Since there is a $\sim 2 \sigma$ inconsistency between the MSAM and the
Saskatoon data and since there may be unknown systematic errors, we have performed
some checks to see how dependent our results are on the various
ways of analyzing the data.

\noindent $\bullet $ Without MSAM\\
The figures and Table I results include the MSAM data points, but
we have also performed $\chi^{2}$ calculations without the MSAM data points. 
When the MSAM  points are not included the $h$ and $\ob$ minimum
and contours do not change significantly.
For example the results from Figure \ref{fig:hob3} without MSAM are
$h=0.30^{+0.18}_{-0.08}$ and $\ob = 0.08^{+0.24}_{-0.08}$.

\noindent  $\bullet$  Saskatoon calibration treatment\\ 
We have treated the Saskatoon calibration as a nuisance parameter from a
Gaussian distribution around the nominal Saskatoon calibration
with a dispersion of $14\%$. The values of $u$ at the minimum $\chi^{2}$
in this technique are $u \sim 0.82$.
Netterfield \etal (1997) have compared the Saskatoon results to the MSAM first flight
results in the north polar region observed by both experiments.
They find a best-fit calibration of $-18\%$ which is equivalent to the $0.82$ discussed
above. Thus, there is some evidence for a lower nominal Saskatoon calibration

However, preliminary results based on a relative calibration between
Jupiter and Cas A at 32 GHz imply that a $+5\% \pm 7\%$ Saskatoon
calibration is appropriate (Leitch \etal 1997).
Using this calibration changes the results slightly.
For example, the analog of Figure \ref{fig:hob3} yields
tighter contours around the unchanged  $h$ minimum:
$h= 0.30 ^{+0.10}_{-0.07}$ and $\ob = 0.19$ with $1\sigma$ error bars
larger than the range probed. The preference for $\ob \sim 0.15 - 0.20$
is increased (independent of $h$) and the avoidance of the high $h$, 
low $\ob$,  BBN region is increased.
The $\chi^{2}$ value of the minimum increases from $\approx 16$ to $\approx 21$
thus the fit is still good; $42\%$ probability of having a lower $\chi^{2}$.

We have also let $u$ be a free parameter from a uniform distribution, i.e.,
a free-floating Saskatoon calibration. The
analog of Figure \ref{fig:hob3} yields $h=0.30^{+0.23}_{-0.07}$,
$\ob = 0.05 ^{+0.20}_{-0.05}$. At the minimum $u=0.78 (=-22\%)$,
$\chi^{2}=14$ and the probability of obtaining a lower $\chi^{2}$ is $7\%$. 
We have also obtained results assuming three different Saskatoon calibrations;
the nominal value, 14\% higher and 14\% lower. The minimum $\chi^2$ stays
at $h \approx 0.30$ in all three cases.

Thus several plausible choices of data selection and data analysis 
producing  $\sim 20\%$ variations in the amplitude of the Doppler peak,
do not strongly affect the low $h$ results from the CMB.
The many measurements on the Doppler slope in the interval $[10 < \ell < 200]$ 
contribute strongly to determining the position of the peak and thus to the 
preference for  low $h$ (see Figures 3 and 4 of paper I).
$h \approx 0.30$ appears to be a fairly robust CMB result for 
the critical density CDM models tested here.

%\clearpage
%%%%%%%%%%%%%%%%%%%%%%%%%%%%%%%%%%%%%%%%%%%%%%%
%\footnotesize
\scriptsize   %referee
%\tiny      %no referee
\begin{table}
\begin{center}
Table I:  {\bf Parameter Results}\\[+2.0mm]
\begin{tabular}{|c l| c c c c|}  \hline 
%        &                                &       &           &       & &\\[-3.0mm]
%\protect\footnotesize
\multicolumn{2}{|c|}{Results$^{a}$} &
\multicolumn{4}{|c|}{ Conditions}\\%   &        
%\multicolumn{1}{|c|}{$\chi^{2}(\%)^{c}$}\\        %P($\;<\chi^{2})^{c}$}  \\
        &                                &    h  &  $\ob(\%)$  & $\n$  &$Q(\mu$K)\\%&        \\
\hline                                                                        
%        &                               &       &             &       &       \\[-2.5mm] % &        \\[-2.5mm]
$H_{o}=$&$35^{+8}_{-5}$                  &   --  & free$^{b}$  &  1    &  17    \\[+1.0mm]%  & 16(11)
        &$35^{+8}_{-5}$                  &   --  & free        &  1    & free   \\[+1.0mm]% & 16(8)  \\[+1.0mm]
        &${\mathbf 30^{+18}_{-7}}$       &   --  & free        & free  & free   \\[+1.0mm]% & 16(9) \\[+1.0mm] 
        &$^{c}40^{+7}_{-7}$              &   --  &   ---       &  --   & --     \\[+1.0mm]% & --     \\[+1.0mm]
 \hline        
$\ob(\%)=$& $2^{+4}_{-2}$                & free  &   --        &  1    &  17     \\[+1.0mm]% & 51(15) \\[+1.0mm]
        & $2^{+4}_{-2}$                  & free  &   --        &  1    & free   \\[+1.0mm] %& 54(15) \\[+1.0mm]
        & ${\mathbf 7^{+24}_{-7}}$       & free  &   --        & free  & free   \\[+1.0mm] %& 55(15) \\[+1.0mm]
        & $^{c}18^{+6}_{-6}$             & --    &   --        & --    &  --    \\[+1.0mm] %& 54(15) \\[+1.0mm]
\hline     
$\n=$   &$1.03^{+0.07}_{-0.06}$          & 0.50  &      5      &   --  & free   \\[+1.0mm] %& 55(15) \\[+1.0mm]
        &${\mathbf 0.91^{+0.20}_{-0.12}}$& free  &    free     &   --  & free   \\[+1.0mm] %& 55(15) \\[+1.0mm]
\hline     
$Q(\mu$K$)=$ & $17^{+1.0}_{-1.0}$                 & free  &    free     &   1   &  --     \\[+1.0mm]% & 55(15)  \\[+1.0mm]
        & $17^{+1.5}_{-2.0}$             & 0.50   &      5      &  free &  --     \\[+1.0mm]%& 55(15)  \\[+1.0mm]
        & ${\mathbf 18^{+2.5}_{-2.5}}$       & free  &     free    &  free &  --    \\[+1.0mm] %& 55(15)  \\[+1.0mm]
\hline
\end{tabular}\\
\end{center}
${}^{a}$ parameter values at the $\chi^{2}$ minimum.
The error bars come from the projected $1\sigma$ contours in the figures.\\

\noindent ${}^{b}$ ``free'' means that the parameters were free to take on any value
within the discretely sampled range which minimized the value of $\chi^{2}$.
As a result of the discretization, these reported minima can be displaced 
from the true minima by up to half a grid point on both axes, i.e., 
$H_{o}=30$ should be taken to mean,
$ 27.5 \le H_{o}  \le 32.5$.
The underlying matrix of model points is described by
$ 0.15 \le h \le 1.00$, step size: 0.05,
$ 0.01 \le \ob \le 0.4$, step size: 0.012,
$ 0.5 \le \n \le 1.5$, step size: 0.06,
$ 10 \le Q \le 30$, step size: 1.
Thus we have tested over 200,000 ($18 \times 34 \times 18 \times 20$) models.\\

\noindent ${}^{c}$ Result from joint likelihood of non-CMB constraints 
(see Figure \protect\ref{fig:hobnocmb}).

\end{table}
\normalsize
%%%%%%%%%%%%%%%%%%%%%%%%%%%%%%%%%%%%%%%%%%%%%%%%%%

\subsection{$\ob$ results}

Our CMB constraints on $\ob$ are weaker than our constraints on $h$; 
the $1\sigma$ contours in Fig \ref{fig:hob3} are elongated vertically and yield
$\ob = 0.07^{+0.24}_{-0.07}$.
Comparing  Figures \ref{fig:hob1} and \ref{fig:hob2} with \ref{fig:hob3},
one sees that it is the marginalization over $\n$ which opens the $\ob \ga 0.07$ 
region (where $n < 0.9$).
White \etal (1996) highlight the merits of high baryonic fraction 
$(\ob \sim 10\% -15\%)$ models.
We confirm that the CMB $\chi^{2}$ $1\sigma$ region is centered near this range 
but the valley of minima is very shallow. 
In the context of our models, non-CMB data can still constrain $\ob$ slightly better than 
the CMB. See our discussion of Figure \ref{fig:hobnocmb} in Sect. \ref{sec:noncmbsummary}
where we report $\ob \approx 0.18 \pm 0.06$.

\section{Non-CMB constraints in the $h - \ob$ plane}
\label{sec:noncmbconstraints}

Four independent non-CMB cosmological measurements constrain the acceptable
regions of the $h - \ob$ plane.

\subsection{Nucleosynthesis}
\label{sec:bbn}
Primordial nucleosynthesis gives us limits on the
baryonic density of the Universe.
Although deuterium measurements seem to be currently the most accurate
baryometer, there is an active debate about whether they yield
high (Tytler \etal 1996, Tytler \& Burles 1997) or  
low  (Songaila \etal 1994, Carswell \etal 1994, Rugers \& Hogan 1996) 
baryonic densities.
We have adopted the range $0.010 < \ob\;h^{2} < 0.026$ because it encompasses
most published results.
These limits are plotted in Figures \ref{fig:hob1} - \ref{fig:hobnocmb} and
are labelled ``BBN'' in Figure \ref{fig:hobnocmb}.
We use $\ob = 0.015$ as a central value.
The BBN constraints are independent of $\oo$, $\ol$, $n$ and $Q$
and thus do not depend on our $\oo = 1$, $\ol = 0$ assumptions.
Lyman limit systems yield a somewhat model-dependent lower limit for
the baryonic density, lending support to the higher values of $\ob h^{2}$
(Weinberg \etal 1997, Bi \& Davidsen 1997). 

\subsection{X-ray cluster baryonic mass fraction}
Observations of the X-ray luminosity and the angular
size of galaxy clusters can be combined to constrain the quantity
$\frac{\ob }{\oo}h^{3/2}$.
We adopt the range $0.04 < \frac{\ob}{\oo} h^{3/2} < 0.10$
(White \etal 1993) with a central value of $0.06$ (Evrard 1997).  
These limits seem to be inconsistent with BBN if 
$\oo = 1$ and $h \ga 0.50$.
This is known as the baryon catastrophy and has led some to believe 
that $\oo < 1$. The severity of this catastrophy in $\oo = 1$ models can be
examined in Figure \ref{fig:hobnocmb} by comparing the ``BBN'' region
with the ``Clusters'' region.
For $\ob \ga 0.15$, $\oo =1$ models allow consistency between the 
nucleosynthetic and cluster data for low values of $h$.

\subsection{Matter power spectrum shape parameter $\Gamma$}
Peacock \& Dodds (1994) made an empirical fit to the
matter power spectrum using a shape parameter $\Gamma$.
For $\oo \le 1$ models, $\Gamma$ can be written as (Sugiyama 1995)
\be
\Gamma = h\:\oo\: exp\left[-\ob\left(\frac{\oo+1}{\oo}\right)\right].
\ee
We adopt the $2\sigma$ limits of the empirical fit of 
Peacock \& Dodds (1994)(see also Liddle \etal 1996a) 
and include the $n$ dependence
\be
\label{eq:gamlimits}
0.222 < \Gamma - 0.32\left(\frac{1}{n} - 1\right) < 0.293.
\ee
Under the assumption that $0.8 \le n \le 1.2$, Equation \ref{eq:gamlimits} becomes
$0.169 \le \Gamma \le 0.373$. 
This is the $\Gamma$ constraint used in Figure \ref{fig:hobnocmb}.
We use $\Gamma = 0.25$ as a central value.

\subsection{Limits on the age of the Universe from the oldest stars in
globular clusters}

There is general agreement that the Universe should be older than the 
oldest stars in our Galaxy. 
Thus a lower bound on the age of the Universe comes
from an age determination of the oldest stars in the most metal-poor
(= oldest) globular clusters.
A reasonably representative 
sample of globular cluster ages, $t_{gc}$, in the literature is,\\

{\baselineskip=.8cm
{\small
\begin{tabular}{llcr}
Bolte \& Hogan (1995)   & $13.7$ &$ < t_{gc} <$& $17.9$\\
Sarajedini \etal (1995) & $13.7$ &$ < t_{gc} <$& $14.3$\\
Chaboyer (1995)         & $11  $ &$ < t_{gc} <$& $21  $\\ 
Jimenez \etal (1996)    & $11.5$ &$ < t_{gc} <$& $15.5$\\
Salaris \etal (1997)    & $10.4$ &$ < t_{gc} <$& $14.0$\\
\end{tabular}
}}\\  %small

\noindent Allowing $\sim 1$ Gyr for the formation of globular clusters,
we adopt the range  $11 < t_{o} < 18$ Gyr with a central value of 14 Gyr.
We use this relatively large interval to avoid overconstraining  the models 
and to encompass most published results.
Age determinations are $\oo$ and $\ol$ independent but converting them
to limits on Hubble's parameter depends on our $\oo=1$ and $\ol=0$ assumptions.
In the models we are considering here, our age limits
are converted directly into limits on Hubble's constant using 
 $h = 6.52\: Gyr/t_{o}$ which yields  $0.59 > h > 0.36$ 
(with the central value 14 Gyr corresponding to $h =0.47$). This region is marked ``Age'' in 
Figure \ref{fig:hobnocmb}.

\subsection{Summary of non-CMB constraints and comparison 
to CMB constraints}
\label{sec:noncmbsummary}

The constraints we adopt from BBN, cluster baryonic fraction, 
$\Gamma$, and the ages of the oldest stars in globular clusters
(as described above) are,\\

{\baselineskip=.8cm
{\small
\begin{tabular}{llcccr}
BBN       &$  0.010  $&$<$&$  \ob\;h^{2}    $&$<$&$ 0.026 $ \\
Clusters  &$  0.04   $&$<$&$  \ob\;h^{3/2}  $&$<$&$ 0.10  $ \\
$\Gamma  $&$  0.169  $&$<$&$   \Gamma       $&$<$&$ 0.373 $ \\ 
GC Ages   &$  0.36   $&$<$&$     h          $&$<$&$ 0.59  $ \\
\end{tabular}
}}\\  %small

\noindent Bands illustrating these constraints are plotted in 
Figure \ref{fig:hobnocmb}.
Since these constraints are independent, it is not obvious that they should be 
consistent with each other. 
They are consistent in the sense that there is a region of overlap.
This consistency is improved if $\ob h^{2}$ turns out to be high
as indicated by Tytler \& Burles (1997).
To visualize more quantitatively the combination of these four constraints,
for each constraint we assume a two-tailed Gaussian distribution
around the central values. 
This allows the flexibility to account for asymmetric error bars.
We then calculate a joint likelihood,

\be
{\mathcal L}(h,\ob)= 
{\mathcal L}_{BBN}*{\mathcal L}_{clus}*{\mathcal L}_{\Gamma}*{\mathcal L}_{age}
\ee
where,
\ba
{\mathcal L}_{BBN}(h,\ob)   &\propto& exp\left[-\frac{(\ob\;h^{2} - 0.015)^{2}}{2\;\sigma_{BBN}^{2}}\right]\\
{\mathcal L}_{clus}(h,\ob)  &\propto& exp\left[-\frac{(\ob\;h^{3/2} - 0.06)^{2}}{2\;\sigma_{clus}^{2}}\right]\\
{\mathcal L}_{\Gamma}(h,\ob)&\propto& exp\left[-\frac{(\Gamma(h,\ob) - 0.25)^{2}}{2\;\sigma_{\Gamma}^{2}}\right]\\
{\mathcal L}_{age}(h)   &\propto& exp\left[-\frac{(h - 0.47)^{2}}{2\;\sigma_{age}^{2}}\right]\\
%{\mathcal L}_{CMB}(h,\ob)   &\propto& exp\left[-\frac{\chi^{2}(h,\ob)}{2}\right].
\ea 

\noindent The upper and lower limits of the four constraints determine the 
$\sigma$'s for the two-tailed Gaussians. For example 
$\sigma_{BBN,up} = (0.026-0.015)$ and
$\sigma_{BBN,down} =(0.015-0.010)$.
The joint likelihood of the four terms
${\mathcal L}_{BBN}*{\mathcal L}_{clus}*{\mathcal L}_{\Gamma}*{\mathcal L}_{age}$ 
is shown in Figure \ref{fig:hobnocmb}. 
The contour levels are ${\mathcal L}(h,\ob)/{\mathcal L}_{max} = exp(-\frac{1}{2}[1,4])$.
The combined non-CMB constraints from BBN, cluster baryonic fraction, 
$\Gamma$ and stellar ages yield
$h \approx 0.40 \pm 0.07$ and $\ob \approx 0.18 \pm 0.06$ for $\oo=1$, $\ol = 0$.

The point of Figure \ref{fig:hobnocmb} {\it is not} to show that $h$ is low since
we have of course ignored the numerous, more direct,  $z \sim 0$ measurements of $h$ 
which find $h \approx 0.70 \pm 0.10$ (see for example Freedman 1997).
The point of this diagram {\it is} to show that an important set of independent constraints 
do overlap and are consistent with each other for low values of $h$
in the $\oo = 1$, $\ol=0$ models considered here.

Figure \ref{fig:hobnocmb} should be compared with Figure \ref{fig:hob3} which has contours of 
${\mathcal L}_{CMB}(h,\ob) \propto exp\left[-\frac{\chi^{2}(h,\ob)}{2}\right]$.
There is an interesting consistency between the 
non-CMB constraints and the CMB constraints of Figure \ref{fig:hob3}.
Although they extend over relatively small regions in the $h - \ob$ plane, 
the $1\sigma$ regions of the non-CMB joint likelihood 
and the CMB overlap. 

In Figure \ref{fig:hobnocmb} we see that the combination of 
four independent cosmological measurements indicate that a low
value of $h$ could make the CDM  theory viable, 
as Bartlett et al. (1995) argued.
The amplitude of small scale matter fluctuations is an additional consistency 
check on this model.
The value of $\sigma_{8}$ at the minimum in Figure \ref{fig:hob3} is
$\sigma_{8} = 0.54$. This agrees quite well with values
inferred from X-ray cluster data (Viana \& Liddle 1996, 
Oukbir \etal 1996).

Liddle \etal (1996b) studied critical density CDM models. Based on the
COBE normalization, peculiar velocity flows, the galaxy correlation function,
abundances of galaxy clusters, quasars and damped Lyman alpha systems, they 
found that $h < 0.50$ and $n < 1$ is preferred.
Adams \etal (1996) come to similar conclusions. 

Bartlett \etal (1995) listed the advantages of a low $h$ in
critical density universes, 
the main point being that there exists a region of
parameter space in which this simplest of models
remains consistent with observations of the large-scale structure of 
the Universe.
For example, there is the question of the age crisis.
Recent $h$ measurements point to values in the interval $ 0.60 < h < 0.80$.
In a critical density universe $h= 0.70$ implies an age of 9.3 Gyr, 
younger than the estimated age of many globular clusters. $h=0.30$ 
yields an age for the universe of 21.7 Gyr
comfortably in accord with globular cluster ages.

What is perhaps surprising is the fact that the
CMB data do not rule out such a such a low value of $H_{o}$
but seem to favor it within the context of this type of scenario.
Of course, these low $H_{o}$ values are in disagreement
with current measurements of the Hubble constant.

%%%%%%%%%%%%%%%%%%%%%%%%%%%%%%%%%%%%%%%%%%%%%%%%%%%%%%
\begin{figure}[htbp]%bp]
%\picplace{5cm}
\centerline{\psfig{figure=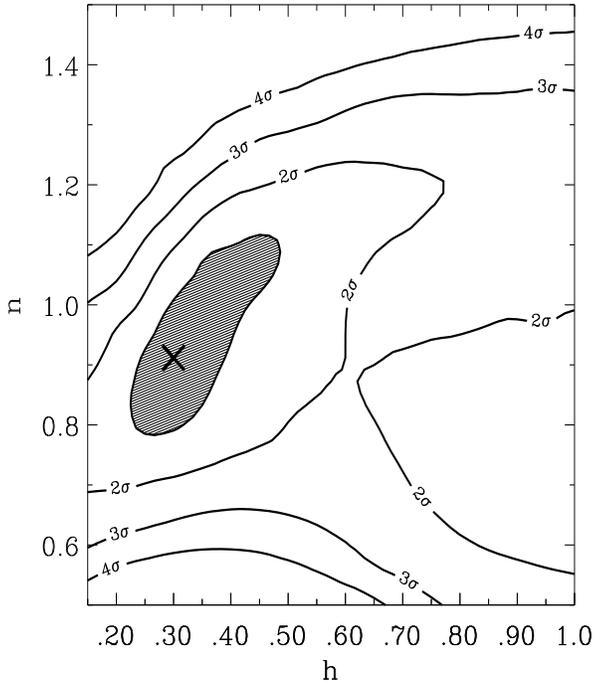,height=10.0cm,width=9.cm,bbllx=10pt,bblly=120pt,bburx=594pt,bbury=690pt}}
%\centerline{\psfig{figure=hn3.ps,height=9cm,width=\hsize,bbllx=10pt,bblly=120pt,bburx=594pt,bbury=690pt}}
%%%BoundingBox: 28 70 594 636
%
\caption{Likelihood contours in the $h - \n$ plane for $\ob$ free and $Q$ free.
$h=0.30^{+0.18}_{-0.07}$ and $n=0.91^{+0.20}_{-0.12}$.
 The $h \protect\ga 0.80$ region has
$Q \protect\sim 20\: \mu$K.
At $2\sigma$, $0.55 < n < 1.23$, and the largest $n$ is for $0.50 < h < 70$.
In the $1\sigma$ contour there is a strong $h$ and $n$ correlation:
the lower values of $h$ go with the lower values of $n$.
When different values of $n$ are allowed,
the amplitude of the Doppler peak varies up and down but
the location of the peak does not.
}
\label{fig:hn3}
\end{figure}
%%%%%%%%%%%%%%%%%%%%%%%%%%%%%%%%%%%%%%%%%%%%%%%%%%%%%%

%%%%%%%%%%%%%%%%%%%%%%%%%%%%%%%%%%%%%%%%%%%%%%%%%%%%%%
\begin{figure}[htbp]%bp]
%\picplace{5cm}
%\centerline{\psfig{figure=hq2n1obfree.ps,height=10.0cm,width=9.cm,bbllx=10pt,bblly=120pt,bburx=594pt,bbury=690pt}}
\centerline{\psfig{figure=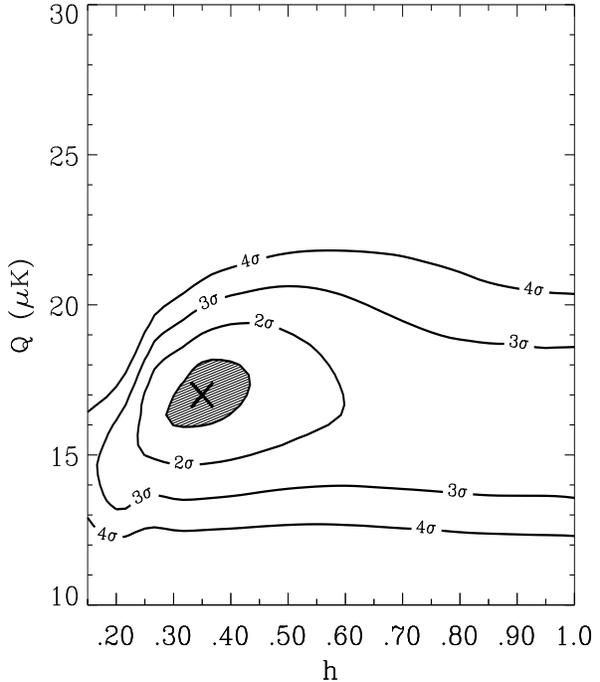,height=10.0cm,width=9.cm,bbllx=10pt,bblly=120pt,bburx=594pt,bbury=690pt}}
%\centerline{\psfig{figure=hq3.ps,height=9cm,width=\hsize,bbllx=10pt,bblly=120pt,bburx=594pt,bbury=690pt}}
%%%BoundingBox: 28 70 594 636
%
\caption{Likelihood contours in the $h - Q$ plane for
$\n=1$ and  $\ob$ free. $h=0.35^{+0.08}_{-0.05}$.
$Q=17 \pm 1.0\;\mu$K.
Notice that since $n$ has been fixed at 1, the value of $Q$ does
not rise in the high $h$ region as it does in the next figure. The
$1\sigma$ contours are round: there is no correlation between $h$ and $Q$.
}
\label{fig:hq2obfree}
\end{figure}
%%%%%%%%%%%%%%%%%%%%%%%%%%%%%%%%%%%%%%%%%%%%%%%%%%%%%%

%%%%%%%%%%%%%%%%%%%%%%%%%%%%%%%%%%%%%%%%%%%%%%%%%%%%%%
\begin{figure}[htbp]%bp]
%\picplace{5cm}
\centerline{\psfig{figure=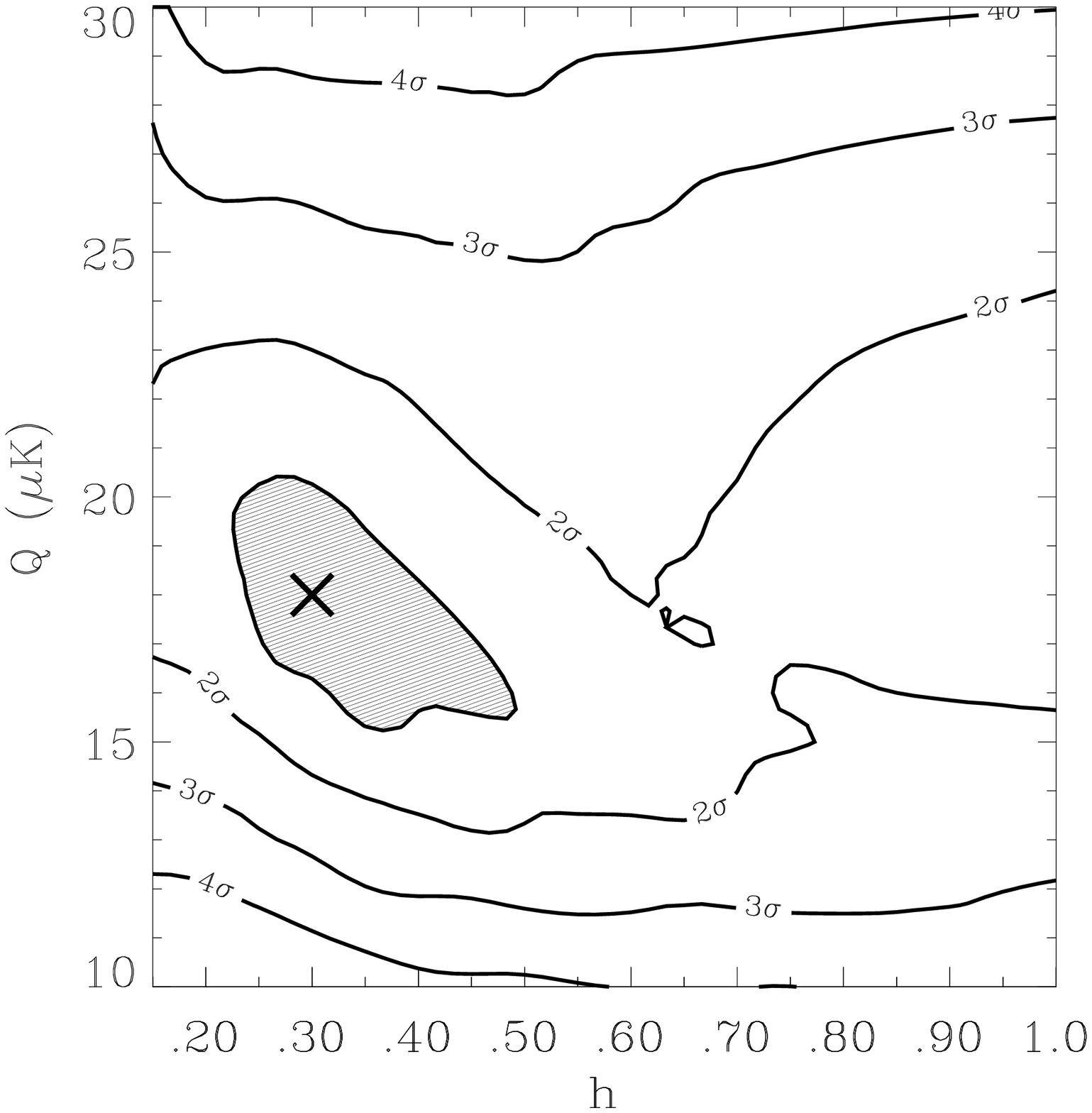,height=10.0cm,width=9.cm,bbllx=10pt,bblly=120pt,bburx=594pt,bbury=690pt}}
%\centerline{\psfig{figure=hq3.ps,height=9cm,width=\hsize,bbllx=10pt,bblly=120pt,bburx=594pt,bbury=690pt}}
%%%BoundingBox: 28 70 594 636
%
\caption{Likelihood contours in the $h - Q$ plane for $\n$ free and $\ob$ free.
Notice that in the high $h$ region, $Q \sim 20\;\mu$K is preferred.
A comparison with Figure \protect\ref{fig:hn3} shows that $n$ and $Q$ are
anti-correlated but this is seen more easily in the next two figures.
}
\label{fig:hq3}
\end{figure}
%%%%%%%%%%%%%%%%%%%%%%%%%%%%%%%%%%%%%%%%%%%%%%%%%%%%%%%%%%%%%%%%%%%%%%%%%%%%%%%

\subsection{Other projections}

\begin{figure}[ht]%bp]
%\picplace{5cm}
\centerline{\psfig{figure=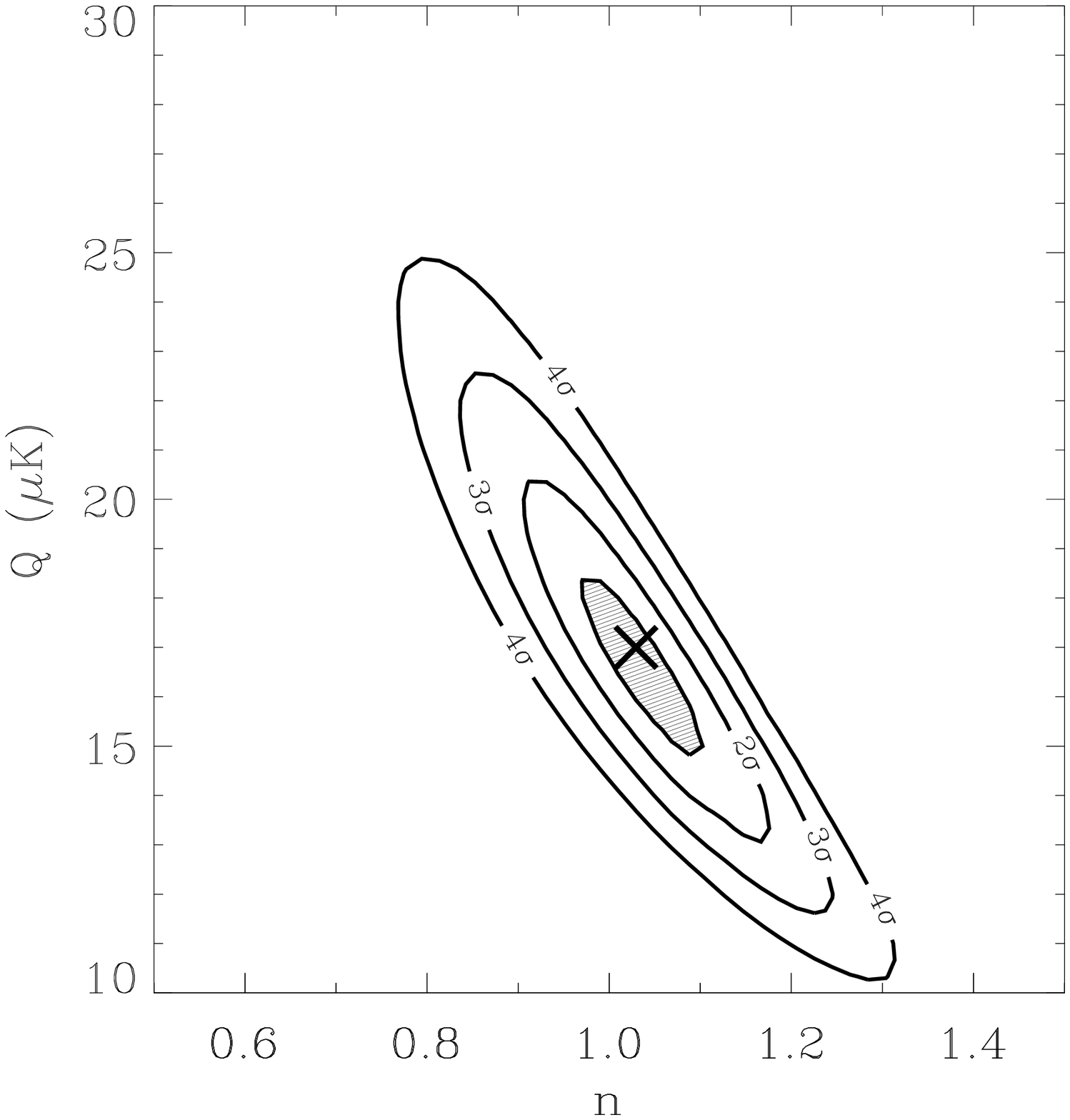,height=10.0cm,width=9.cm,bbllx=10pt,bblly=120pt,bburx=594pt,bbury=690pt}}
%\centerline{\psfig{figure=nq1h50ob5.ps,height=10.0cm,width=9.cm,bbllx=10pt,bblly=120pt,bburx=594pt,bbury=690pt}}
%\centerline{\psfig{figure=nq1.ps,height=9cm,width=\hsize,bbllx=10pt,bblly=120pt,bburx=594pt,bbury=690pt}}
%%%BoundingBox: 28 70 594 636
%
\caption{ Likelihood contours in the $n - Q$ plane for $h=0.50$ and $\ob = 0.05$.
Result $n=1.03^{+0.07}_{-0.06}$ and $Q=17^{+1.5}_{-2.0}$,
or at the 95\% CL, $0.90 \leq n \leq 1.17$ and
$13 \leq Q \leq 20.5\;\mu$K.
This standard model has been looked at by several other workers
who obtain similar results (see Sect. \protect\ref{sec:nqresults}).
}
\label{fig:nq1}
\end{figure}

\begin{figure}[ht]%tbp]
%\picplace{5cm}
\centerline{\psfig{figure=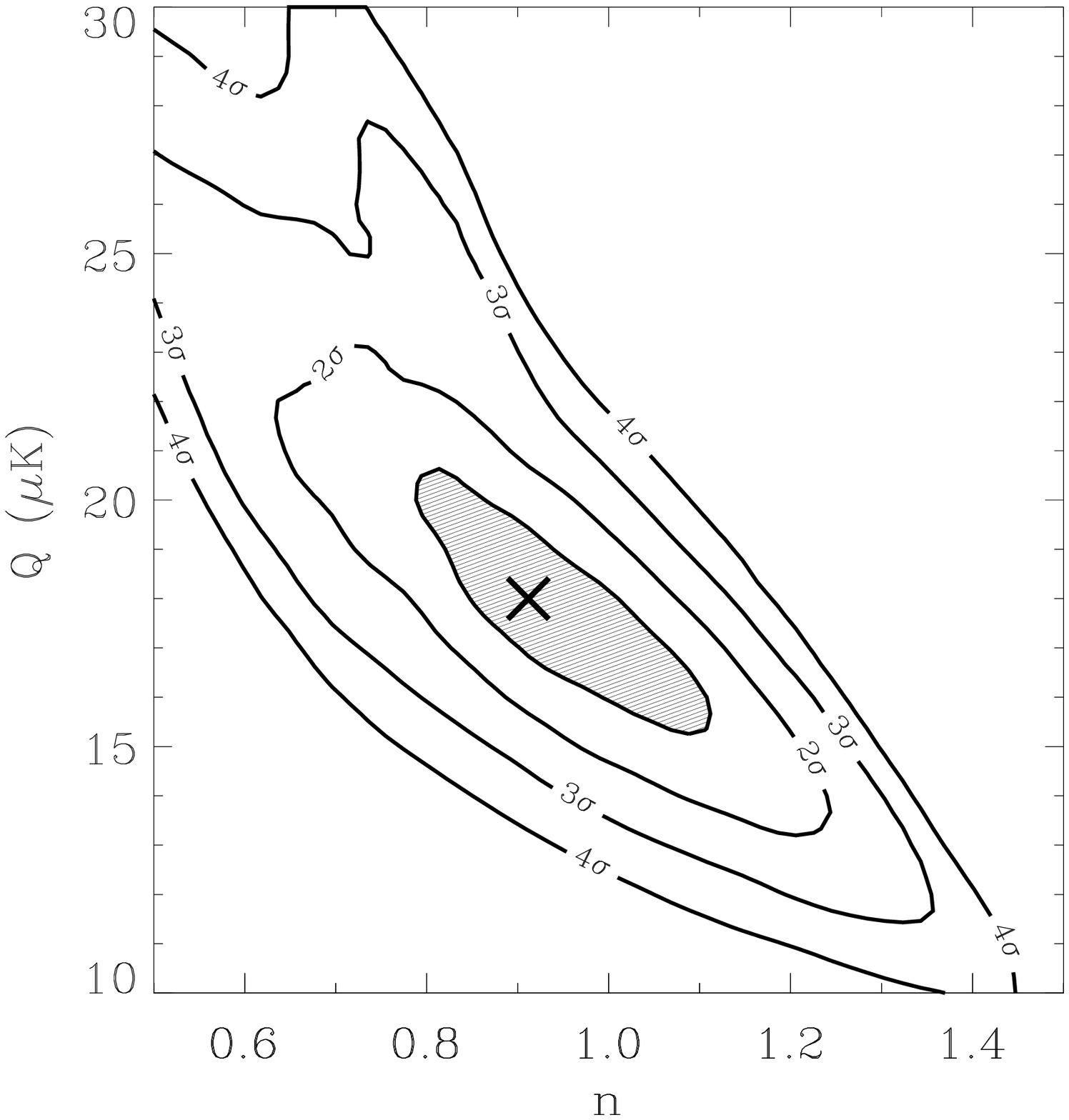,height=10.0cm,width=9.cm,bbllx=10pt,bblly=120pt,bburx=594pt,bbury=690pt}}
%\centerline{\psfig{figure=nq3.ps,height=10.0cm,width=9.cm,bbllx=10pt,bblly=120pt,bburx=594pt,bbury=690pt}}
%\centerline{\psfig{figure=nq3.ps,height=9cm,width=\hsize,bbllx=10pt,bblly=120pt,bburx=594pt,bbury=690pt}}
%%%BoundingBox: 28 70 594 636
%
\caption{Likelihood contours in the $n - Q$ plane for $h$ free and $\ob$ free.
$n=0.91^{+0.20}_{-0.12}$, $Q=18 \pm 2.5\;\mu$K.
The minimum is at $h=0.30$, $\ob = 0.07$, and $u=0.82$.
}
\label{fig:nq3}
\end{figure}

Figures \ref{fig:hn3}  through \ref{fig:nq3}
show our 4-D $\chi^{2}$ ellipsoid projected on to 2-D planes orthogonal to the $h-\ob$ plane.
The limits obtained on $h$ from Figure \ref{fig:hn3} are the same as we obtained from
Figure \ref{fig:hob3} since we are projecting the same 4-D $\chi^{2}$ surface.
$h$ and $n$ are positively
correlated for $h \la 0.50$ and possibly negatively correlated
for $h \ga 0.50$. 
In Figure \ref{fig:hq2obfree} we see that with $n$ fixed at 1, a high
precision determination of $Q$ is possible, $Q=17\pm 1.0\;\mu$K.
In Figure \ref{fig:hq3} the $n=1$ constraint is dropped.

%%%%%%%%%%%%%%%%%%%%%%%%%%%%%%%%%%%%%%%%%%%%%%%%%%%%%%
%%%%%%%%%%%%%%%%%%%%%%%%%%%%%%%%%%%%%%%%%%%%%%%%%%%%%%

%%%%%%%%%%%%%%%%%%%%%%%%%%%%%%%%%%%%%%%%%%%%%%%%%%%%%%
%%%%%%%%%%%%%%%%%%%%%%%%%%%%%%%%%%%%%%%%%%%%%%%%%%%%%%%%%%%%%%%%%%%%%%%%%%%%%%%%%%

\section{Results  for $n$ and $Q$}
\label{sec:nqresults}
Figures \ref{fig:nq1} and \ref{fig:nq3} show the strong anti-correlation between $\n$ 
and $Q$. This has been observed and discussed
by many authors (e.g. Smoot \etal 1992, Seljak \& Bertschinger 1993, Lineweaver 1994).
This anti-correlation has a simple explanation.
In the $C_{\ell}(\ell)$ plot, any increase of the slope lowers the
y-intercept (at $\ell = 2$) and any decrease of the slope raises the 
y-intercept. The anti-correlation is thus inherent with the use of the amplitude
at $\ell = 2$ as the normalization.

The increase of the size of the error bars on $n$ and $Q$ as $h$ and $\ob$ are 
conditioned on and then marginalized can be seen by comparing Figures \ref{fig:nq1} 
and \ref{fig:nq3}.
Our $n$ and $Q$ results are $n= 0.91^{+0.20}_{-0.12}$ and
$Q=18 \pm 2.5\:\mu$K from Figure \ref{fig:nq3} where both $h$ and $\ob$ have been 
marginalized.
Conditioning on $n=1$, we get $Q=17 \pm 1.0\; \mu$K (see also Fig. \ref{fig:hq2obfree}).

The four year COBE-DMR constraints on the amplitude and slope of the power spectrum at 
large angular scales are $\n_{DMR}=1.2 \pm 0.3$ and $Q=15.3^{+3.8}_{-2.8}\:\mu$K, and
conditioning on $n_{DMR}=1$, $Q=18 \pm 1.6\: \mu$K (Bennett \etal 1996).
These DMR results are in the context of $\oo = 1$, $\ol = 0$ CDM models
and they need to be corrected due to the mildly model-dependent
extended tails of the Doppler peak which reach even into the low-$\ell$ region. 
After the correction, the DMR result becomes
$\n\approx 1.05 \pm 0.3$ and $Q \approx 17.5^{+3.8}_{-2.8}\:\mu$K.
Thus our results from a combination of recent CMB measurements in the context of 
critical density universes agrees well with the DMR-only result and reduces the error
bars on both $n$ and $Q$.

For the standard CDM model of Figure \ref{fig:nq1} 
($h=0.50$, $\oo=1$ and $\ob \approx 0.05$),
we obtain $0.90 \leq n \leq 1.17\:(95\%$ CL) and $13 \leq Q \leq 20.5\;\mu$K (95\% CL).
Using similar methods and a similar data set, several authors have 
reported similar results.
de Bernardis \etal (1996) find $1.0 \leq n \leq 1.26\: (95\%$ CL).
White \etal (1996) find $0.86 < n < 1.10$ ($95\%$ CL).
Hancock \etal (1997) find $1.0 < n < 1.2$ ($68\%$ CL).
The variations of these standard model $n$ determinations are probably due to 
slightly different data sets, different treatments of the Saskatoon calibration
and the details of the $\chi^{2}$ calculation. 

\section{Caveats and summary}
\label{sec:summary}

\subsection{Caveats}
\label{sec:caveats}

Our low $h$ measurements are in disagreement with current $z \sim 0$  measurements 
of the Hubble constant. Possible explanations for this discrepancy could be
unidentified systematics in the CMB data or the local $h$ measurements.
Galactic foregrounds could be a problem for the CMB while the distance ladder
may need some readjusting for the local $h$ measurements
(e.g., Feast \& Catchpole 1997).
The best way to address these problems is with more and better data.
This is being done rapidly. New detectors and better designed observations
are improving the quality of both the CMB data and the more direct $h$ measurements.
SZ and supernovae observations are also increasing the redshift of the
$h$ measurements. 

A simple answer to the discrepancy between direct $h$  measurements and 
our results is that the Universe is not well-described by the models
considered here. It is possible that one or more of our basic assumptions is wrong,
or we could simply be looking at too restricted a region of parameter space. 
The shape of the primordial power spectrum may 
be more complicated than the family of models we are using.
Inflation may be wrong and structure may not have formed from
adiabatic curvature fluctuations.
Topological defects may be the origin of structure.

We emphasize that we have only considered
a particular type of cosmological scenario, although arguably the simplest; 
the results we have presented here are valid under the assumption
of inflation-based, Gaussian adiabatic initial conditions in a critical 
density universe ($\oo =1$) with no cosmological constant. 
We have not considered any early reionization scenarios or gravitational wave
contribution. We have also not included any hot dark matter.

Our $\oo = 1$ assumption can be considered very restrictive since plausible values for
$\oo$ in the range $[0.2, 1.0]$ can change the power spectrum significantly.
For example, the position of the Doppler peak, $\ell_{peak}$ is roughly proportional to
$\oo^{-1/2}$. Thus low $\oo$, by pushing $\ell_{peak}$ higher
may permit higher $h$ values to accomodate the high $\ell_{peak} \sim 270$ of the
CMB data. We are in the process of checking this assumption;
we consider open models in Lineweaver \etal (1997, in preparation) 

Reionization models can affect the power spectra significantly
by lowering the Doppler peak but this can be compensated by $n$ values
larger than 1. For example, deBernardis \etal (1996) have looked at 
reionization models and find a best-fit early reionization at $z_{reion} \approx 20$
and $n\approx 1.2$.

In paper I we took a brief look at flat-$\lambda$ models.
The supernovae results of Perlmutter \etal (1997) can constrain $\lambda$
better than the CMB data. The combination of the CMB, BBN and supernovae
constraints in flat-$\lambda$ models yields $0.23 < h < 0.72$.

If gravitational waves or any other effect plays an important role in CMB anisotropy
formation, we expect that the inclusion of this effect in the family of models 
tested, will improve the resulting fits. However the inclusion of gravitational waves seems
to make the fits slightly worse without changing the
location of the minimum (Liddle \etal 1996b).
Bond and Jaffe (1997) analyzed the combined DMR (Bennett \etal 1996), 
South Pole (Gunderson \etal 1995) and Saskatoon (Netterfield \etal 1997)
data using signal-to-noise eigenmodes. They looked at the parameters
$h$, $n$ and $\sigma_{8}$ in a variety of models. 
The inclusion of tensor modes for $n < 1$ models
seems to produce small shifts in the likelihood surfaces.

There may be extra-relativistic degrees of freedom (hot dark matter (HDM) or
mixed dark matter (MDM)).
deBernardis \etal (1996) found that
current CMB anisotrophy measurements cannot distinguish between CDM and MDM models.
We agree with this assessment and add that HDM and CDM models also cannot usefully be 
distinguished with current CMB data.

In addition to the  $h$, $\ob$, $n$ and  $Q$ considered here,
regions of a larger dimensional parameter space deserve further investigation
including $\oo$, $\ol$, $\oh$, early reionization parameters such as $z_{reion}$, 
tensor mode parameters $n_{T}$ and $T$, iso-curvature or adiabatic initial 
conditions and topological defect models with their additional parameters such as the 
coherence length. 
The fact that we obtain acceptable $\chi^{2}$ values in our small 4-D parameter
space lends some support to the idea that we may be close to the right model.
However establishing error bars on broad-band power estimates is a relatively new science.
If the Universe is not well-described by these models then as the data improve,
work like this will show poor $\chi^2$ fits and other regions of parameter space may
be preferred. 

\subsection{Summary}

CMB measurements have become sensitive enough to constrain cosmological 
parameters in a restricted class of models.
The results we have presented here are valid under the assumption
of Gaussian adiabatic initial conditions in a critical 
density ($\oo =1$) universe  with no cosmological constant. 
We have explored the 4-dimensional parameter space of $h$, $\ob$, $n$ and $Q$.
Our CMB-derived  constraints on $h$, $\ob$, $n$ and $Q$ exclude 
large regions of parameter space.
 We obtain a low value for Hubble's constant:
$H_{o}= 30 ^{+18}_{-7}$. 
The CMB data constrain $\ob$ only weakly: $\ob = 0.07^{+ 0.24}_{-0.07}$.
For the slope and normalization of the
power spectrum we obtain  $\n=0.91^{+0.20}_{-0.12}$ and $Q=18 \pm 2.5\:\mu$K.
The error bars on each parameter are
for the case where the other 3 parameters have been marginalized.
When  we condition on $n=1$ we obtain the normalization $Q= 17 \pm 1.0\:\mu$K.

CMB constraints are independent of other cosmological tests
of these parameters and are thus particularly important. 
The fact that reasonable $\chi^2$ values are obtained means that
the current CMB data are consistent with inflationary-based
CDM models with a low Hubble constant. 
In the context of the models considered, the CMB results are consistent with
four other independent cosmological measurements but
are in disagreement with more direct measurements of $h$.

\section{Acknowledgements}
The rapidly increasing quality and quantity of data along with the fast 
Boltzmann code developed by Uros Seljak and Matias Zaldarriaga has 
made this work possible.
We thank Alain Blanchard and Jim Bartlett for useful discussion. 
We thank Martin White, Douglas Scott and the MAX group for help
assembling the required experimental window functions.
C.H.L. acknowledges
support from the French Minist\`ere des Affaires Etrang\`eres
and NSF/NATO post-doctoral fellowship 9552722.
D.B. is supported by the Praxis XXI CIENCIA-BD/2790/93 grant attributed by 
JNICT, Portugal.

%**********************************************************************
%\clearpage
%%%%%%%%%%%%%%%%%%%%%%%%%%%%%%%%%%%%%%%%%%%%%%%%%%%%%%%%%%%%%%%%%%%
%\clearpage

%**********************************************************************

\end{document}